\documentclass[12pt,preprint]{aastex}
\usepackage{longtable}
\usepackage{epstopdf}
\usepackage{rotating}
\usepackage{graphicx}
\usepackage{float}
\usepackage[flushleft]{threeparttable}
\usepackage{amsmath}
\newcommand{\beq}{\begin{equation}}
\newcommand{\eeq}{\end{equation}}
\newcommand{\beqn}{\begin{eqnarray}}
\newcommand{\eeqn}{\end{eqnarray}}

\begin{document}
 
\title{\bf{Are There Transit Timing Variations for the Exoplanet Qatar-1b ?}
}
\author{Li-Hsin Su$^1$, Ing-Guey Jiang$^1$, Devesh P. Sariya$^1$, 
Chiao-Yu Lee$^1$,  Li-Chin Yeh$^2$,
Vineet Kumar Mannaday$^3$, Parijat Thakur$^3$,
D. K. Sahu$^4$, Swadesh Chand$^3$, 
A. A. Shlyapnikov$^5$,
V. V. Moskvin$^5$, 
Vladimir Ignatov$^5$,
David Mkrtichian$^6$, Evgeny Griv$^7$}  

\affil{
{$^1$Department of Physics and Institute of Astronomy,}\\ 
{National Tsing-Hua University, Hsin-Chu, Taiwan}\\
{email: jiang@phys.nthu.edu.tw}\\
{$^2$Institute of Computational and Modeling Science,}\\
{National Tsing-Hua University, Hsin-Chu, Taiwan}\\
{$^3$Department of Pure \& Applied Physics,}\\
{Guru Ghasidas Vishwavidyalaya (A
Central University), Bilaspur (C.G.)-495009, India}\\
{$^4$Indian Institute of Astrophysics, Bangalore-560034, India}\\
{$^5$Crimean Astrophysical Observatory, 298409, Nauchny, Crimea}\\
{$^6$National Astronomical Research Institute of Thailand (NARIT),}\\
{Siripanich Building, 191 Huaykaew Road, Muang District, Chiangmai, Thailand}\\
{$^7$Department of Physics, Ben-Gurion University, Beer-Sheva 84105, Israel}
}
  
\begin{abstract}
Motivated by the unsettled conclusion on whether there are any 
transit timing variations (TTVs) for the exoplanet Qatar-1b,
10 new transit light curves are presented and the TTV analysis
with a baseline of 1400 epochs are performed.
Because the linear model provides a good fitting
with reduced chi-square $\chi^{2}_{red} = 2.59$ and 
the false-alarm probabilities of possible TTV frequencies 
are as large as 35\%,
our results are consistent with a null-TTV model.
Nevertheless, a new ephemeris with the reference time 
$T_0 =2455647.63360\pm 0.00008$ (BJD) and 
the period 
$P= 1.4200236\pm 0.0000001$ (day) 
is obtained.
In addition, the updated 
orbital semi-major axis and planetary radius in unit of stellar radius
are being provided, and the lower limit of 
modified stellar tidal quality factor is also determined.
\end{abstract}
 
\keywords{planetary systems $-$
stars: individual: Qatar-1 $-$ transit $-$ techniques: photometric}
 
\section{Introduction}
\label{Intro}

The increasing focus on the science of extrasolar planets (exoplanets) 
is one of the most prominent features of astrophysics
in the 21$^{st}$ century.  
Thousands of extrasolar planets have been discovered and 
the main credit goes to the methods of Doppler shift and transits. 
While the Doppler-shift detection technique played 
a major role in the initial phase,
the transit method has played a more 
vital role in finding new planetary systems in recent years.
The new transit discoveries caused an unprecedented jump in the number 
of known exoplanets owing to the satellite observations by
CoRoT (Baglin et al. 2006), Kepler (Borucki et al 2010), 
and the updated version of Kepler, i.e. K2 mission (Howell et al. 2014). 
However, the role of ground-based observations has been very crucial as well.
Various ground-based surveys such as TrES (Alonso et al. 2004), 
SuperWASP (Pollacco et al. 2006), 
KELT (Pepper et al. 2007), MASCARA (Talens et al. 2017),  
Qatar (Alsubai et al. 2013) etc. 
have discovered many exoplanets. 
 
Hot Jupiters, the preferred targets for the ground-based transit surveys,
are the gas giants found at closer orbital distances
with masses larger than $0.25 M_J$ and radii about 1 or 2 $R_J$ 
(Labadie-Bartz et al. 2018). Even though hot Jupiters are rare 
according to the occurrence rate (Dawson \& Johnson 2018), 
the sensitivity of current observing techniques favors their detection,
as they have deep transits ($\sim$1\%) 
and short orbital periods (1 to 10 days) which enable multiple
observations at a short interval to confirm their planetary nature 
(Hellier et al. 2018). 

While the total number of known exoplanets has been increasing steadily, 
the above exciting results have triggered many theoretical investigations
and statistical studies. For example, the planet formation has been modeled
and addressed in Mordasini et al. 2009). The orbital evolution 
has been studied in Jiang \& Ip (2001), Ji et al. (2002),
Jiang et al. (2003), Jiang \& Yeh (2004a, 2004b, 2007),  
Gayon \& Bois (2008). The distributions of exoplanets on 
period-mass plane were addressed in  
Zucker \& Mazeh (2002), Tabachnik \& Tremaine (2002), 
and Jiang et al. (2006).
Additionally, the coupled period-mass functions were first 
explored in Jiang et al. (2007, 2009), and then further investigated
with proper treatments of the selection effect in Jiang et al. (2010).
Moreover, Jiang et al. (2015) studied the 
period-ratio-mass-ratio correlation
of adjacent planet pairs in 166 multiple planetary systems.
A moderate correlation between the period-ratio and mass-ratio
was found with a correlation coefficient 0.5303.

In addition to the above theoretical and statistical studies,
the transit observations of known planetary systems also lead to 
new implications.  
When the periodicity of the transit timing is not a constant, 
it is related to the transiting exoplanet orbiting in a 
non-Keplerian potential which could be caused by the presence 
of additional planets in the system (Linial, Gilbaum \& Sari 2018). 
The deviations from a linear ephemeris is called transit timing variations 
(TTVs). In a known exoplanetary system, there can be some undiscovered planets.
The hindrance in their detection is caused by the limitations of 
detection techniques and sensitivity of available instruments. 
In such cases, the TTVs can play a very crucial role.
Thus, the in-depth follow-up studies of these systems are needed for
the study of transit timing variations 
and the characterization of planetary systems. 

In recent years, many TTV studies have been carried out. For example,
Maciejewski et al. (2010) showed that
a periodical TTV was confirmed and was likely due to an additional 15 
Earth-Mass planet orbiting near the outer
2:1 mean-motion resonance in the WASP-3 system. 
The later work in
Maciejewski et al. (2013) did not validate the existence of that 15 Earth-Mass 
planet, but determined the upper limit on the mass of any hypothetical 
additional planet.
For the TrES-3 system, though Lee et al. (2011) favored
a linear fit with a constant period,
Jiang et al. (2013) reported possible TTVs with their
five new transit light curves.  Recently, Mannaday et al. (2020)
revisited this system with new transit observations of later epochs
and confirmed the possibility of TTVs. 
 
In addition to giving implications of the presence of unknown planets,
TTVs could be an indication of the orbital decay of the transit planet. 
For example, through the photometric monitoring on WASP-43b, 
Blecic et al. (2014) claimed an orbital period decreasing rate  
about 0.095 second per year. Similarly, 
Murgas et al. (2014) proposed an orbital decay
with period decreasing rate of about 0.15 second per year.
With eight new transit light curves, 
Jiang et al. (2016) gave a new measurement on the rate of orbital decay 
which is one order of magnitude smaller than the previous values.
This new result of a slow decay led to a normally assumed theoretical
value of stellar tidal dissipation factor, and thus resolve the 
previous controversial situation.
Later, Hoyer et al. (2016) presented a result with an even smaller 
decay rate. Finally, Chernov et al. (2017) compared their
theoretical results with the observational decay rates in 
both Jiang et al. (2016) and Hoyer et al. (2016) and claimed that 
their theory of dynamical tides can explain these observations
when all theoretical regimes are considered.

These interesting results attracted a lot of attention from 
the astronomy community 
and triggered many further investigation on the orbital decay of hot Jupiters.
For example, Maciejewski et al. (2016) worked on WASP-12b and
claimed that there is likely an orbital decay with a rate about 
0.0256 sec per year. 
In addition, Patra et al. (2017) employed both the transit and 
occultation times to further constrain the orbit of WASP-12b.
They considered both the apsidal precession and orbital decay models,
and concluded that the orbital decay was more likely. 
Recently, Baluev et al. (2019) also gave a confirmation on 
the orbital decay of WASP-12b.
 
Among many discovered planetary systems, Qatar-1 is the one which has been
investigated with unsettled conclusions.
Qatar-1 is an old star with an age larger than 4 Gyr.
It is a metal-rich dwarf star with K3 spectral type.  
Around this star, there is one confirmed planet which was discovered by
Qatar Exoplanet Survey (Alsubai et al. 2011). 
The planet, Qatar-1b, is a hot Jupiter with an orbital period 
of $\sim$1.42 days. 
The planet's mass is $\sim$1.275 $M_J$ and its radius is $\sim$1.136 $R_J$.  
It moves on a circular orbit with an orbital inclination $i_b= 84^\circ$.26, 
and semi-major axis $a_b$/$R_*$= 6.319 (Maciejewski et al. 2015).
Covino et al. (2013) derived more accurate orbital parameters 
and spin-orbit alignment for the system.
They also found the Qatar-1 to be a chromospherically active star. 
The first TTV analysis of the system was carried out by von Essen et al. (2013).
They claimed that there are possible TTVs for Qatar-1b and 
speculated that the 190-day TTVs can be caused by 
a weak perturbator in resonance with Qatar-1b or by a massive body 
similar to a brown dwarf.
However, the follow-up TTV studies by Maciejewski et al. (2015)
and Collins et al. (2017) did not find 
any evidence of additional planet in the system. 
In contrast, P\"usk\"ull\"u et al. (2017) found weak evidence of
TTVs based on their later transit timing analysis. 

Thus, Qatar-1b is an interesting target for further study.
The controversial situation motivated us 
to obtain new transit light curves in order to investigate this system. 
Employing both our new light curves and published data,
we will be able to cover more epochs than the previous work.

More information about the data used in this work 
can be seen in Section~\ref{OBS}, where we also describe 
the selection of comparison stars.
The analysis of light curves is presented in Section~\ref{LC}.
The comparison of this work with the previous studies is 
presented in Section~\ref{PREVIOUS}, followed by
the TTV analysis with related frequencies and models in Section~\ref{TTV}.
Finally, the conclusions are provided in Section~\ref{CON}. 

\section{Observational Data}
\label{OBS}

In this paper, we employed our own newly observed data in 
combination with many published light curves from multiple sources.
All of them are homogeneously normalized through the same procedure.

\subsection{Observations}

Three telescopes at different sites were employed 
for our transit observations. 
One is the 60-inch telescope (P60) of Palomar Observatory
located at Palomar mountain in north San Diego County, California, USA. 
The observing facility belongs to California Institute of Technology. 
P60 is a reflecting telescope built with Ritchey-Chretien optics, 
and both primary and secondary mirrors have hyperbolic reflection surface. 
The mean seeing quality of images is about 1.9 arcsec. 
The optical imager uses a 2048$\times$2048-pixel$^2$ CCD array camera 
to image roughly 13$\times13$ arcmin$^2$ of sky.

Another telescope we used is the 0.5-meter telescope, MTM-500, which belongs to
the Crimean Astrophysical Observatory (CrAO) at Nauchny in Crimea 
(Longitude $34^\circ$1$\arcmin$ E, Latitude $44^\circ$32$\arcmin$ N). 
The camera mounted on MTM-500 is Apogee Alta U6 and the CCD array 
contains 1024$\times$1024-pixels$^2$. The field of view is about 
12 $\times$ 12 arcmin$^2$, 
and the image seeing is about 5$\arcsec$ (0.71$\arcsec$/pixel plate scale). 

In addition,
we also observed with the 2-meter Himalayan Chandra Telescope (HCT), 
located at Hanle India. 
The observations of the Qatar-1 system were taken in 
the Bessell R-filter by gathering 60 seconds exposures
using the Himalaya Faint Object Spectrograph Camera (HFOSC), 
equipped with a 2048 $\times$ 4096 pixels$^2$ CCD 
with a pixel size of 15 microns. 
The central 2048 $\times$ 2048 region of the CCD used for 
imaging covers a field of view of 10 $\times$ 10 arcmin$^2$ on the sky, 
with a scale of 0.296 arcsec pixel$^{ -1}$.
These three HCT light curves were presented by some of us in a 
workshop\footnote[1]{The First Belgo-Indian Network for 
Astronomy \& Astrophysics (BINA) Workshop, November 2016, 
Nainital, India}  
(Thakur et al. 2018).

In this work, we have used six light curves from P60, 
one light curve from MTM-500 and three light curves from HCT.
They are all complete transit light curves without any obvious interruption.
The details of the observation log of these data are presented 
in Table 1. 
Since we set the zero epoch for one of the transits taken from
von Essen et al. (2013), the transit epoch for Run 1 is 891.  

%Table 1
\begin{table}
\centering
\label{log}
\begin{tabular}{ccccccc}
\hline\hline
Run   & Epoch & UT Date  & Telescope  & Filter & Exposure Time(s) & PNR($\%$)\\
\hline
1 &891  & 2014 Sep. 12 &1.5-m P60       & R & 20 & 0.23\\
2 &898  & 2014 Sep. 22 &1.5-m P60       & R & 20 & 0.21\\
3 &905  & 2014 Oct. 02 &1.5-m P60       & R & 20 & 0.25\\
4 &911  & 2014 Oct. 10 &0.5-m MTM-500   & R & 60 & 0.55\\
5 &924  & 2014 Oct. 29 &1.5-m P60       & R & 20 & 0.22\\
6 &1034 & 2015 Apr. 03 &1.5-m P60       & R & 20 & 0.22\\
7 &1053 & 2015 Apr. 30 &1.5-m P60       & R & 20 & 0.34\\
8 &1354 & 2016 Jun. 30 &2.0-m  HCT      & R & 60 & 0.22\\
9 &1361 & 2016 Jul. 10 &2.0-m  HCT      & R & 60 & 0.16\\
10&1399 & 2016 Sep. 02 &2.0-m  HCT      & R & 60 & 0.19\\
\hline
\end{tabular}
\caption{The log of observations of this work. 
From Run 1 to Run 10, the transit epoch, 
the UT date, telescope, filter, exposure time,
and photometric noise rate are shown.
}
\end{table}

\subsection{The Selection of Comparison Stars} 

We adopted a method following the procedure in Jiang et al. (2016) 
to select suitable comparison stars. First of all, we expect that the stars 
lying near the target and having similar magnitudes would be 
good choices to be the comparison stars. In other words, we rule out 
those stars which are not around the target, 
or those which have much brighter or fainter magnitudes.
Among these selected candidates, the brightness consistency is checked
in order to ensure none of them are variable objects. 
Then, the flux of the target is divided by any possible combination 
of the fluxes of these candidate stars. 
This leads to many calibrated light curves, and the one with
the smallest standard deviation for the out-of-transit part
becomes the target's light curve.
Those candidate stars which contribute to the target's light curve
are termed as comparison stars.

\subsection{Other Data}

The published Qatar-1b transit light curves were searched. 
We employed two selection criteria to choose the appropriate light curves,
The first criterion was that there should be no obvious interruption 
in the light curves and the second was that there should be data points 
on both sides of out-of-transit parts. 
We included 28 published light curves into our analysis: 
9 light curves from von Essen et al. (2013), 
5 light curves from Covino et al. (2013), 
and 14 light curves from Maciejewski et al. (2015).

\subsection{The Time Stamp}

We used the Barycentric Julian Date 
in Barycentric Dynamical Time (BJD) 
as the time stamps in all light curves.
The Universal Time (UT) is obtained from the recorded header.
We then compute the UT of mid exposure and convert
the time stamp to BJD (Eastman et al. 2010).
The numerical data of these 10 new light curves are in Table 2.

%Table2 
\begin{table}
\centering
\begin{tabular}{lcccc}
\hline\hline
Run & Epoch & BJD & Relative Flux & Uncertainty\\
\hline
1  & 891      & 2456912.821530 & 0.998609 & 0.000920\\
   &          & 2456912.822064 & 0.996934 & 0.000918\\
   &          & 2456912.822591 & 1.003170 & 0.000924\\
   &          & 2456912.823118 & 1.000120 & 0.000921\\ 
   &          & 2456912.823644 & 1.001340 & 0.000922\\
   &          &         -      &      -   &     -   \\
\hline
2   & 898      & 2456922.760684 & 1.001020 & 0.000922\\ 
    &          & 2456922.761213 & 1.006300 & 0.000927\\ 
    &          & 2456922.761759 & 1.007900 & 0.000928\\
    &          & 2456922.762289 & 1.005690 & 0.000926\\
    &          &2456922.7628180 & 1.004180 & 0.000925\\
    &          &         -      &      -   &     -   \\
\hline
3   & 905     & 2456932.704141  & 1.000160 & 0.000921\\ 
    &         & 2456932.704674  & 1.002360 & 0.000923\\ 
    &         & 2456932.705207  & 0.999670 & 0.000921\\
    &         & 2456932.705740  & 1.002750 & 0.000924\\ 
    &         & 2456932.706272  & 1.001150 & 0.000922\\   
    &         &        -        &    -     &     -   \\
\hline
4   & 911      & 2456941.193138 & 1.009980 & 0.003721\\
    &          & 2456941.194551 & 1.005870 & 0.003706\\
    &          & 2456941.195268 & 0.998864 & 0.003680\\
    &          & 2456941.195974 & 1.007950 & 0.003713\\ 
    &          & 2456941.196692 & 0.997118 & 0.003673\\ 
    &          &     -          &    -     &    -    \\
\hline
 -  &   -  &     -             &     -     &    -    \\
\hline  
\end{tabular}
\caption{The data of our photometric light curves. 
This table is available in its entirety in machine-readable forms.
}
\end{table}

%Table 3
\begin{table}
\centering
\begin{tabular}{ccc}
\hline\hline
Parameter & Initial Value & During MCMC Chain \\
\hline
$P$ (day) & 1.42002406   & a Gaussian prior with $\sigma = 0.00000021$, not linked\\
$i$ (deg) & 84.26        & free, linked \\
$a/R_{*}$  & 6.319        & free, linked \\
$R_{p}/R_{*}$ & 0.14591      & free, linked \\
$T_m$        & set by TAP   & free, not linked \\
$u_1$        & 0.55790921   & a Gaussian prior with $\sigma = 0.05$, not linked\\
$u_2$        & 0.16000079   & a Gaussian penalty with $\sigma = 0.05$, not linked\\
$e$          & 0.0          & fixed\\
$\omega$     & 0.0          & fixed\\
\hline
\end{tabular}
\caption{The parameter setting. 
The initial values of $P$ (day), $i$ (deg), $a/R_{*}$, and $R_{p}/R_{*}$
are taken from Maciejewski et al. (2015).
}
\end{table}

\section{Transit Analysis}
\label{LC}

In order to determine the mid-transit times and important 
planetary parameters from light curve data, 
the Transit Analysis Package (TAP), developed by Gazak et al. (2012),
is employed in the present study. 
It is an interface-driven software package designed 
for the analysis of exoplanet transit light curves. 
TAP software uses Markov Chain Monte Carlo (MCMC) 
technique to fit light curves by using the model of Mandel \& Agol (2002) 
and the wavelet-based likelihood function developed by Carter \& Winn (2009). 
There are nine parameters to describe the planetary system: 
orbital period $P$, 
scaled semi-major axis $a/R_{*}$, 
scaled planet radius $R_{p}/R_{*}$, 
orbital inclination $i$, 
mid-transit time $T_{m}$, 
the linear limb darkening coefficient $u_1$, 
the quadratic limb darkening coefficient $u_{2}$, 
orbital eccentricity $e$, and longitude of periastron $\omega$.\\

All 38 light curves are loaded into the TAP. 
Among these light curves, 10 light curves are obtained by our own observations 
and 28 light curves are from published literature. 
For each light curve, five MCMC chains of 300,000 steps are computed.  
To start a MCMC chain in TAP, 
initial values of the above parameters are needed. 
The initial value of $P$, $i$, $a/R_{*}$, and $R_{p}/R_{*}$
are all taken as the values in Maciejewski et al. (2015), i.e.
$P=1.42002406$,  $i$=84.26,  $a/R_{*}=6.319$, and $R_{p}/R_{*}=0.14591$.
The initial value of $T_{m}$ is set by TAP automatically.

The linear and quadratic limb darkening parameters were calculated by the 
EXOFAST package (Eastman et al. 2013) which interpolates 
limb darkening tables in Claret $\&$ Bloemen (2011).
With stellar parameters such as metallicity $[Fe/H]=0.2\pm$0.1, 
surface gravity log $g=4.55\pm0.1$ (in CGS unit),
and effective temperature $T_{eff}=4910\pm100K$
(Covino et al. 2013, Maciejewski et al. 2015), 
we obtained the linear and quadratic limb darkening 
as 0.56 and 0.16, respectively.  
Moreover, $e$ and $\omega$ are both set to zero, 
based on Alsubai et al. (2011). 
They compared the results of circular and eccentric orbits and 
found the circular orbit to be more reliable. 

With the above initial values, during MCMC chains,
parameters can be  
(1) completely fixed, (2) completed free to vary, or 
(3) varying with a Gaussian prior.
When Option (2) or (3) is chosen for a particular parameter, 
one can further choose whether this parameter is linked among all
light curves or not. This linking means all light curves are 
taken into account while this parameter is determined through 
data-model fitting during MCMC chains, and thus only one value 
would be obtained for this parameter. 
Three parameters $i$,  $a/R_{*}$, and $R_{p}/R_{*}$, 
are linked during TAP runs.

The period $P$ is allowed to vary following a Gaussian function 
with $\sigma = 0.00000021$ (day). 
The choice of $\sigma$ of Gaussian priors for period $P$ 
is based on the corresponding uncertainties of $P$ 
in Maciejewski et al. (2015). 
No Gaussian priors are set
for the parameters $i$, $a/R_{*}$, and $R_{p}/R_{*}$.
In addition, the mid-transit time  $T_m$ is completely free during TAP runs. 
The limb darkening coefficients are allowed 
to vary around the theoretical values following a Gaussian function 
with $\sigma = 0.05$ (e.g.  M{\"u}ller et al. 2013). 
Finally, 
orbital eccentricity $e$ and longitude of periastron $\omega$ are fixed 
so that they will not change during the MCMC process.  
The detailed information of parameter setting for TAP runs are 
summarized in Table 3. 

The transit models for all light curves are obtained after TAP runs.
We show 10 new light curves and the corresponding residuals  
of our own observations in Figure 1. 
The points are observation data and the solid lines are models.
Following Fulton et al. (2011), the photometric noise rates (PNR)
of these light curves are determined and listed in Table 1.
In addition, Table 4 presents the list of resulting mid-transit times.   

\section{The Comparison with Previous Work}
\label{PREVIOUS}

In order to demonstrate that our analysis procedure 
is consistent with the ones used in literature,
we compared our results for the published light curves with the values
mentioned in the corresponding papers.
The comparison was done with the light curves taken from
von Essen et al. (2013),  Covino et al. (2013), and 
Maciejewski et al. (2015). 

At first, nine mid-transit times of Data Source (a),
five mid-transit times of Data Source (b),
14 mid-transit times of Data Source (c) 
are taken from Table 4. 
They are our TAP results and are denoted as $T_{m1}$ here.
These $T_{m1}$ are used to do a linear fitting
with a straight line defined as: 
\begin{equation}
C_m(E)= T_0 + P E,
\end{equation}
where $P$ is the period, $E$ is an epoch, and $T_{0}$
is a reference time. 
By minimizing $\chi^2$ through the above linear fitting,
we obtain 
$P = 1.4200238\pm 0.0000002$ (day) and
$T_0 = 2455647.63352^{+0.00010}_{-0.00009}$ (BJD).
The value of reduced chi-square is $\chi^{2}_{red} = 3.09$. 
Note that all minimum $\chi^2$ fittings in this paper 
were performed through the MCMC sampler called {\it emcee}
(Foreman-Mackey et al. 2013).

For the comparison purpose, we took the corresponding values 
of published mid-transit times, denoted as $T_{m2}$, 
and error bars from von Essen et al. (2013), 
Covino et al. (2013), and Maciejewski et al. (2015),  
for the same light curves. 
By minimizing $\chi^2$ through the same linear fitting,
we obtain 
$P = 1.4200240\pm 0.0000002$ (day) and
$T_0 =2455647.63339\pm 0.00012$ (BJD). 
The reduced chi-square is $\chi^{2}_{red} = 1.72$. 

When the above mid-transit times are subtracted by the 
corresponding values of $C_m$, the value $T_{m1}-C_m$ or $T_{m2}-C_m$ 
as a function 
of epoch $E$ provides the $O-C$ plot. 
Figure 2 is the $O-C$ plot, where green points represent 
results from our TAP runs,
while red points show the published mid-transit times.
For most of the points, green error bars overlap with red error bars,
which shows that the results are consistent.

In order to further quantify the differences between
our re-determined mid-transit times and the published mid-transit times,
we define
\begin{equation}
\Delta T = \frac{T_{m1} - T_{m2}}{\sqrt{(\sigma_1^2 + \sigma_2^2)/2}} , 
\end{equation}
where $\sigma_1$ is the error bar of $T_{m1}$ and
$\sigma_2$ is the error bar of $T_{m2}$.
Figure 3 presents the resulting $\Delta T$ as a function of epoch. 
It is clear that the values of $\Delta T$ are all within the range [-1,1].
This means that the differences between $T_{m1}$ and $T_{m2}$
are around the order of their error bars. 
Thus, our re-determined mid-transit times are confirmed to be 
consistent with the published mid-transit times.

%Table 4
\begin{table}
\centering
\begin{tabular}{ccc}
\hline\hline
Epoch & Data source & $T_m$ (BJD-2455000) \\
\hline
0    & (a) & $647.63228^{+0.00031}_{-0.00033}$\\ 
45   & (b) & $711.53484^{+0.00019}_{-0.00021}$\\
90   & (c) & $775.43472^{+0.00047}_{-0.00048}$\\
107  & (b) & $799.57628^{+0.00018}_{-0.00017}$\\
133  & (a) & $836.49650^{+0.00025}_{-0.00025}$\\
238  & (a) & $985.60066^{+0.00060}_{-0.00064}$\\
283  & (a) & $1049.50000^{+0.00033}_{-0.00035}$\\
290  & (a) & $1059.43980^{+0.00110}_{-0.00114}$ \\
302  & (a) & $1076.47944^{+0.00065}_{-0.00065}$ \\
328  & (a) & $1113.39955^{+0.00084}_{-0.00080}$ \\
340  & (b) & $1130.44153^{+0.00022}_{-0.00022}$ \\
347  & (a) & $1140.38099^{+0.00034}_{-0.00033}$ \\
359  & (a) & $1157.42050^{+0.00051}_{-0.00054}$ \\
364  & (b) & $1164.52243^{+0.00019}_{-0.00020}$ \\
376  & (b) & $1181.56266^{+0.00014}_{-0.00014}$ \\
442  & (c) & $1275.28513^{+0.00027}_{-0.00029}$ \\
628  & (c) & $1539.40769^{+0.00029}_{-0.00029}$ \\
647  & (c) & $1566.38856^{+0.00030}_{-0.00032}$ \\
704  & (c) & $1647.33079^{+0.00031}_{-0.00029}$ \\
776  & (c) & $1749.57123^{+0.00051}_{-0.00050}$ \\
807  & (c) & $1793.59250^{+0.00038}_{-0.00037}$ \\
814  & (c) & $1803.53275^{+0.00018}_{-0.00019}$ \\
833  & (c) & $1830.51335^{+0.00018}_{-0.00018}$ \\
840  & (c) & $1840.45409^{+0.00038}_{-0.00039}$ \\
890  & (c) & $1911.45493^{+0.00020}_{-0.00018}$ \\
891  & (d) & $1912.87445^{+0.00039}_{-0.00038}$ \\
895  & (c) & $1918.55483^{+0.00033}_{-0.00034}$ \\
898  & (d) & $1922.81460^{+0.00030}_{-0.00028}$ \\
902  & (c) & $1928.49500^{+0.00030}_{-0.00028}$ \\
905  & (d) & $1932.75501^{+0.00035}_{-0.00035}$ \\
911  & (d) & $1941.27534^{+0.00099}_{-0.00088}$ \\
923  & (c) & $1958.31411^{+0.00082}_{-0.00084}$ \\
924  & (d) & $1959.73497^{+0.00042}_{-0.00046}$ \\
1034 & (d) & $2115.93766^{+0.00028}_{-0.00028}$ \\
1053 & (d) & $2142.91872^{+0.00044}_{-0.00043}$ \\
1354 & (d) & $2570.34627^{+0.00032}_{-0.00032}$ \\
1361 & (d) & $2580.28552^{+0.00014}_{-0.00014}$ \\
1399 & (d) & $2634.24662^{+0.00022}_{-0.00022}$ \\
\hline
\end{tabular}
\caption[NewTable4]{The results of light-curve analysis for the  
mid-transit time $T_m$. 
Data sources: (a) von Essen et al. (2013), (b) Covino et al. (2013), 
(c) Maciejewski et al. (2015), and (d) this work.}
\end{table}

%Table 5
\begin{table}
\label{midtrans}
\centering
\begin{tabular}{ccccc}
\hline\hline
Parameter & $P$ & $i$ & $a/R_{\star}$ & $R_{p}/R_{\star}$  \\
This work & $1.4200236\pm 0.0000001$ & $84.26^{+0.11}_{-0.10}$ &
$6.33^{+0.05}_{-0.04}$ & $0.1456^{+0.0005}_{-0.0006}$ \\ 
Covino et al. (2013) & $1.42002504\pm 0.00000071$ & 
$83.82\pm 0.25$ & $6.25\pm 0.10$ &
$0.1513\pm 0.0008$ \\
von Esson et al. (2013) & $1.4200246\pm 0.0000004$ &
$84.52\pm 0.24$ & $6.24\pm 0.10$ &
$0.1435\pm 0.0008$ \\
Maciejewski et al. (2015) & $1.42002406 \pm 0.00000021$ &
$84.26^{+0.17}_{-0.16}$ & $6.319^{+0.070}_{-0.068}$ &
$0.14591^{+0.00076}_{-0.00078}$\\
\hline
\end{tabular}
\caption[NewTable5]{The redetermined system parameters, 
together with the values from literature.}
\end{table}

% ^{+0.25}_{-0.25}  ^{+0.10}_{ -0.10} ^{+0.0008}_{-0.0008}

\section{Transit Timing Variations}
\label{TTV}

\subsection{New Ephemeris}

The new ephemeris is determined by minimizing the $\chi^{2}$ through 
fitting observational mid-transit times to a linear function,
i.e. Eq. (1).
We obtained
$T_0 =2455647.63360\pm 0.00008$ (BJD),
$P= 1.4200236\pm 0.0000001$ (day), 
and the corresponding $\chi^{2} = 93.34$. Because the degree of freedom is 36, 
the reduced chi-square would be $\chi^{2}=2.59$.
Using this new ephemeris, the $O-C$ plot is presented in  Figure 4. 
This value of $P$ is the newly determined orbital period in this work.
It is listed in Table 5 in order to have a direct comparison 
with those results in previous work.
In addition, the values of $i$, $a/R_{*}$, and $R_{p}/R_{*}$  are
also presented in Table 5. 

\subsection{The Frequency Analysis}

In order to examine whether there are any periodical variations 
in the $O-C$ diagram,
the generalized Lomb-Scargle periodogram (Zechmeister \& Kuerster 2009),
which takes the effect of error bars into account,    
was used to search for possible frequencies in the data. 
The periodogram is a plot with the spectral power as a function of frequency
as shown in Figure 5. 
The highest peak of spectral power is at frequency $f=0.01544$ (epoch$^{-1}$).

In order to quantitatively determine how 
significant this possible frequency is,
the false-alarm probability (FAP) can be employed.
For a given frequency with spectral power $z$, 
the false-alarm probability $P(z)$,
which determines if a given frequency is not a real frequency,
can be calculated through 
the empirical bootstrap resampling algorithm.

It is found that the FAP of this frequency is $35\%$. 
It means that the probability for
this frequency to be a real periodicity
is close to $65\%$. This is too low to be a significant signal.
The result of our frequency analysis indicates that there is 
no periodic TTV for Qatar-1b.

\subsection{The Rate of Orbital Decay}

As shown in Jiang et al. (2016), Hoyer et al. (2016), 
and Wilkins et al. (2017),
the TTV analysis can give a good constraint on the rate
of orbital decay. Moreover, this rate could lead to a measurement
of modified stellar tidal quality factor. 

Following Wilkins et al. (2017) and Maciejewski et al. (2018),
the mid-transit times of  
a model of orbital decay can be expressed as
\begin{equation}
T_m(E)=T_0 + P E + \frac{1}{2}\frac{d P}{d E} E^2,
\end{equation}
where $T_0$ is a reference time, $P$ is the orbital period, $E$ is an epoch, 
and $dP/dE$ is the change in the orbital period between 
succeeding transits.
Fitting the above model with observational mid-transit times 
through $\chi^2$ minimization,
we find that  
$T_0 = 2455647.63350 \pm 0.00012$ (BJD),
$P = 1.4200240 \pm 0.0000004$ (day), 
$dP/dE = (-5.9\pm 5.2)\times {10^{-10}}$,
and the corresponding $\chi^{2} = 92.01$.
Because the degree of freedom is 35, the 
$\chi^{2}_{red} = 2.63$. 
It can be noticed that the error bar of $dP/dE$ is the same order
as the best-fit value, and $dP/dE$ is effectively around zero.

Moreover, from
Wilkins et al. (2017) and Maciejewski et al. (2018), 
the ratio of stellar tidal quality factor to 
the second-order stellar tidal Love number $k_2$, i.e. 
the modified stellar tidal quality factor $Q^{'}_{\star}$, 
can be expressed as
\begin{equation}
Q^{'}_{\star} = - \frac{27}{2}\pi \frac{M_p}{M_{\star}}
(\frac{R_{\star}}{a})^5 (\frac{dP}{dE})^{-1} P,
\end{equation}
where 
$M_{p}$ is the planet mass, $M_{\star}$ is the stellar mass, 
$R_{\star}$ is the stellar radius,
$a$ is the semi-major axis, 
and $P$ is the period. 
With the values of $M_{p}$ and $M_{\star}$ taken from Collins et al. (2017),
the values of $P$ and $a/R_{\star}$ from Table 5,   
and the 5th percentile of the posterior distribution of $dP/dE$,
we obtain a lower limit for $Q^{'}_{\star}$ as
$Q^{'}_{\star}  >   8.8 \times 10^6$.

\section{Conclusions}
\label{CON}

After being discovered by Qatar Exoplanet Survey (Alsubai et al. 2011),
the exoplanet Qatar-1b was further studied by various groups photometrically.
von Essen et al. (2013) suggested a possible 190-day TTVs
which can be produced by a third body. 
While Maciejewski et al. (2015) and Collins et al. (2017) did not find 
any TTVs in this system, 
P\"usk\"ull\"u et al. (2017) found a weak indication of 
TTVs based on their later analysis. 

In order to solve this controversial situation, we employed several 
telescopes to monitor the Qatar-1 system and 
obtained 10 new transit light curves.
Combining with another 28   
published light curves, we performed a TTV analysis that covers
a baseline of 1400 transit epochs. 

Firstly, through a linear fitting, the new ephemeris was 
established with the reference time 
$T_0 =2455647.63360\pm 0.00008$ (BJD) and 
the period $P= 1.4200236\pm 0.0000001$ (day). 
This fitting leads to a reduced chi-square $\chi^{2}_{red}=2.59$.

Secondly, the generalized Lomb-Scargle algorithm
was used to search for possible TTV frequencies. 
It was determined that the false-alarm probability is 
$35\%$ for the identified frequency, and thus no significant
periodicity is found.

Given that the reduced chi-square $\chi^{2}_{red}=2.59$ for 
the linear fitting and no TTV frequencies are identified,
we conclude that our result is consistent with a null-TTV
model for the exoplanet Qatar-1b.
Nevertheless,
taking the advantage of this TTV analysis, the lower limit of 
the modified stellar tidal quality factor is  
determined as $Q^{'}_{\star}  > 8.8 \times 10^6$.
In addition, the orbital inclination, 
the orbital semi-major axis and planetary radius 
in unit of stellar radius 
are updated.

\section*{Acknowledgment}
We are thankful to the referee for very helpful suggestions. 
This work is supported by the grant from
the Ministry of Science and Technology (MOST), Taiwan.
The grant numbers are
MOST 105-2119-M-007 -029 -MY3
and MOST 106-2112-M-007 -006 -MY3. 
PT and VKM acknowledge the University Grants
Commission (UGC), New Delhi, for providing the financial
support through Major Research Project no. UGC-MRP 43-
521/2014(SR).
PT and VKM thank the staff at IAO, Hanle and
CREST (IIA), Hosakote for providing support during the observations. The
time allocation committees of the HCT is also gratefully acknowledged for
providing the observation times.
PT expresses his sincere thanks to IUCCA, Pune, India for providing the
supports through IUCCA Associateship Programme.

\clearpage
%Fig1: 10 light curves
\begin{figure*}
\begin{center}
\label{residual}
\centering
\includegraphics[scale=0.7]{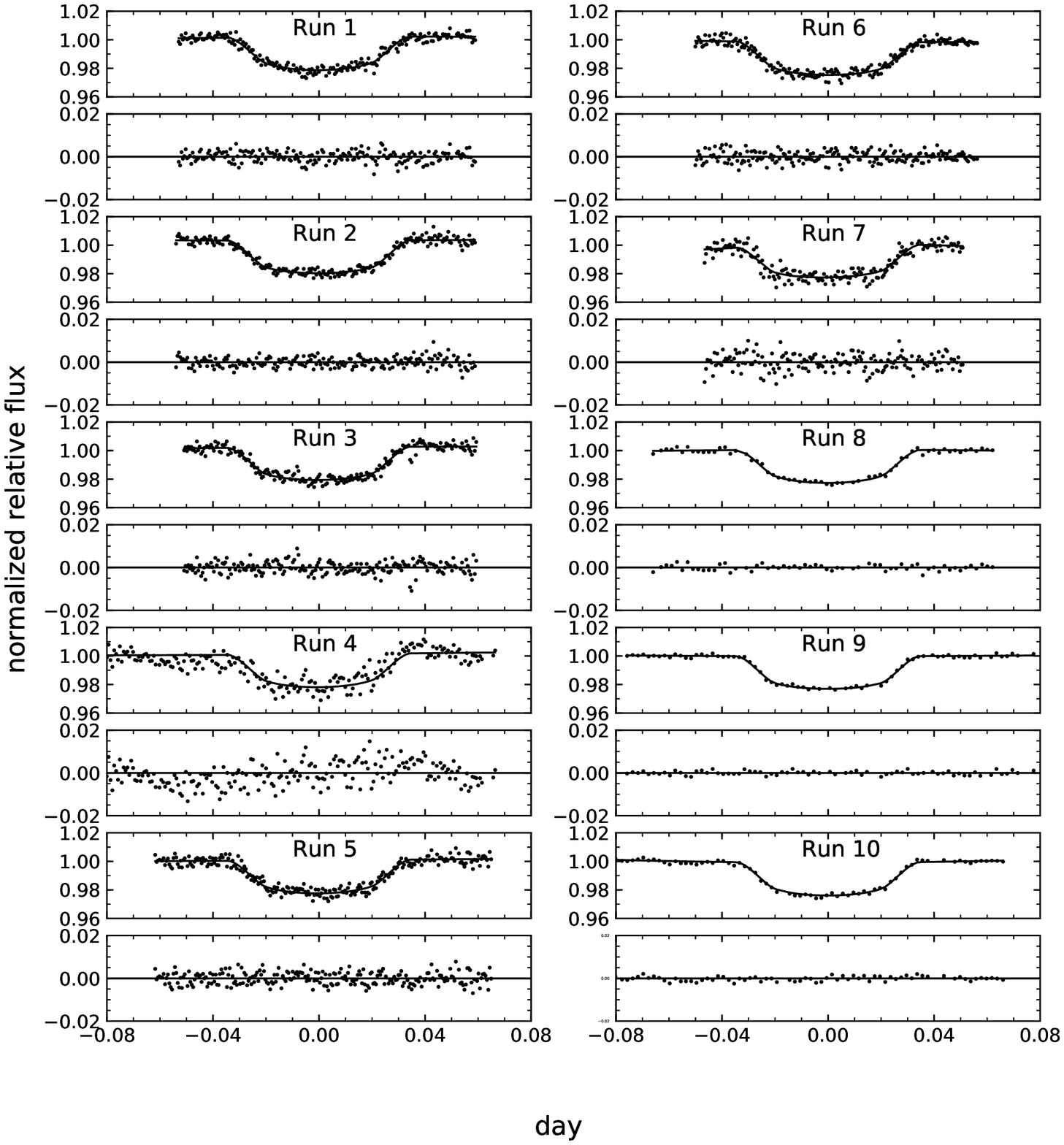}
\caption{The normalized relative flux as a function of time 
from Run 1 to Run 10. The points are data and the solid lines are models.
The corresponding residuals are shown under each light curve. 
The unit for $x$-axis is day 
(offset from mid-transit time and in TDB-based BJD).
}
\end{center}
\end{figure*}

%Fig2
%%%%%%%%%%%%%%%%%%% Comparision O-C %%%%%%%%%%%%%%%%%%%%%%%%%%%%
\begin{figure*}
\begin{center}
\label{O-Cvonreference}
\includegraphics[scale=1]{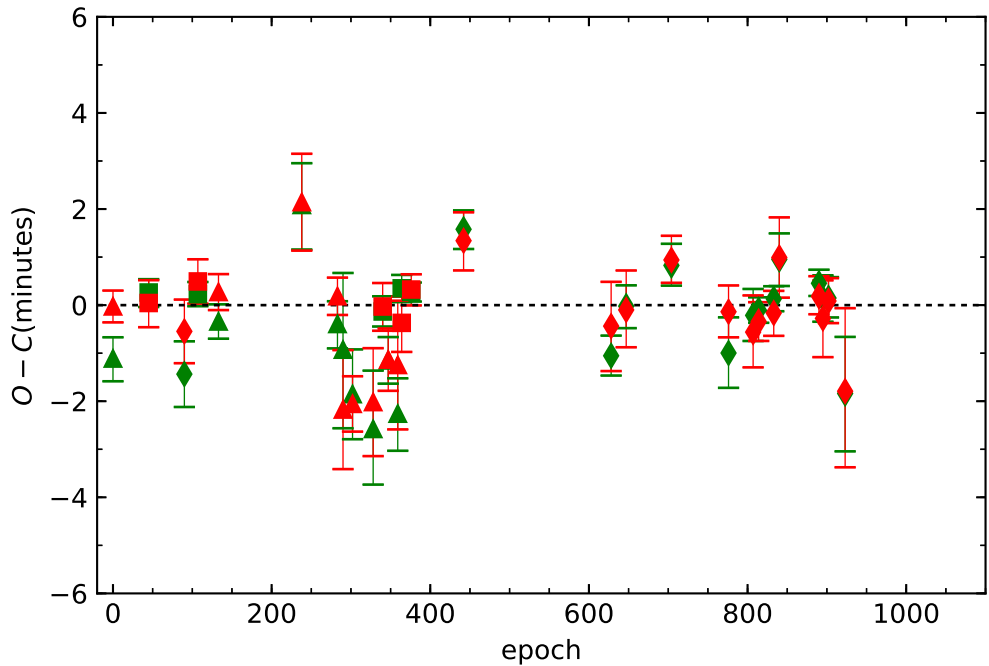}
\caption{The $O-C$ diagram, i.e. the deviation of mid-transit times 
from the fitted ephemeris as a function of epoch.
The green points are derived from our TAP runs and 
the red points are derived from previously published mid-transit times.
Triangles are the results derived from light curves in von Essen et al. (2013).
Squares are the results derived from light curves in Covino et al. (2013). 
Diamonds are the results derived from light curves in 
Maciejewski et al. (2015). 
}
\end{center}
\end{figure*}

%Fig 3
%%%%%%%%%%%%%%%%%%% Delta T %%%%%%%%%%%%%%%%%%%%%%%%%%%%
\begin{figure*}
\begin{center}
\label{O-CCovinoreference}
\includegraphics[scale=1]{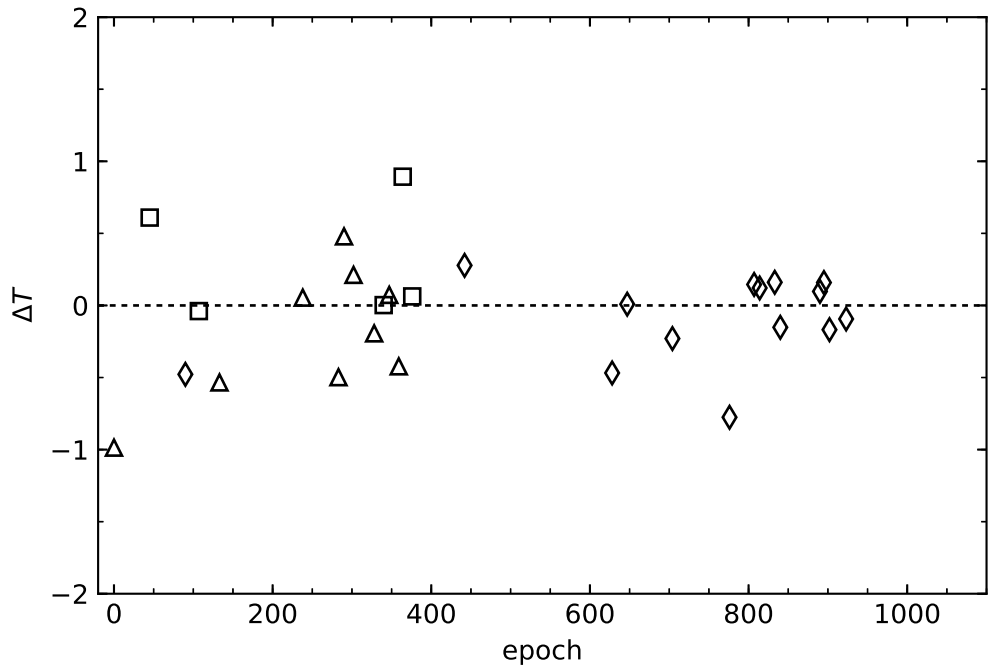}
\caption{The value of $\Delta T$ as a function of epoch.
Triangles are the results derived from light curves in von Essen et al. (2013),
squares are the results derived from light curves in Covino et al. (2013),
and diamonds are derived from light curves in 
Maciejewski et al. (2015). 
}
\end{center}
\end{figure*}

%Fig 4
%%%%%%%%%%%%%%%%%%%%%%% O-C %%%%%%%%%%%%%%%%
\begin{figure*}
\begin{center}
\label{O-Clinear}
\includegraphics[scale=1]{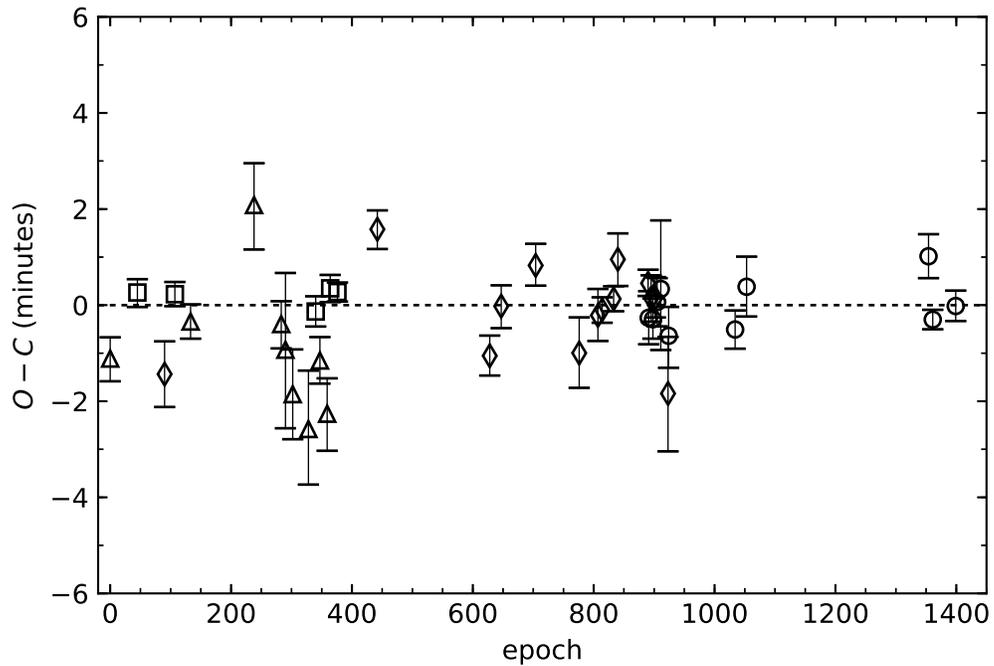}
\caption{The $O-C$ diagram, i.e. the deviation of mid-transit times 
from the new ephemeris as a function of epoch, 
for the results of all employed light curves.
Triangles are the results derived from light curves in von Essen et al. (2013).
Squares are the results derived from light curves in Covino et al. (2013). 
Diamonds are the results derived from light curves in 
Maciejewski et al. (2015). 
Circles are the results of our own observational data.}
\end{center}
\end{figure*}

%Fig 5  spectral power
\begin{figure*}
\begin{center}
\includegraphics[scale=1]{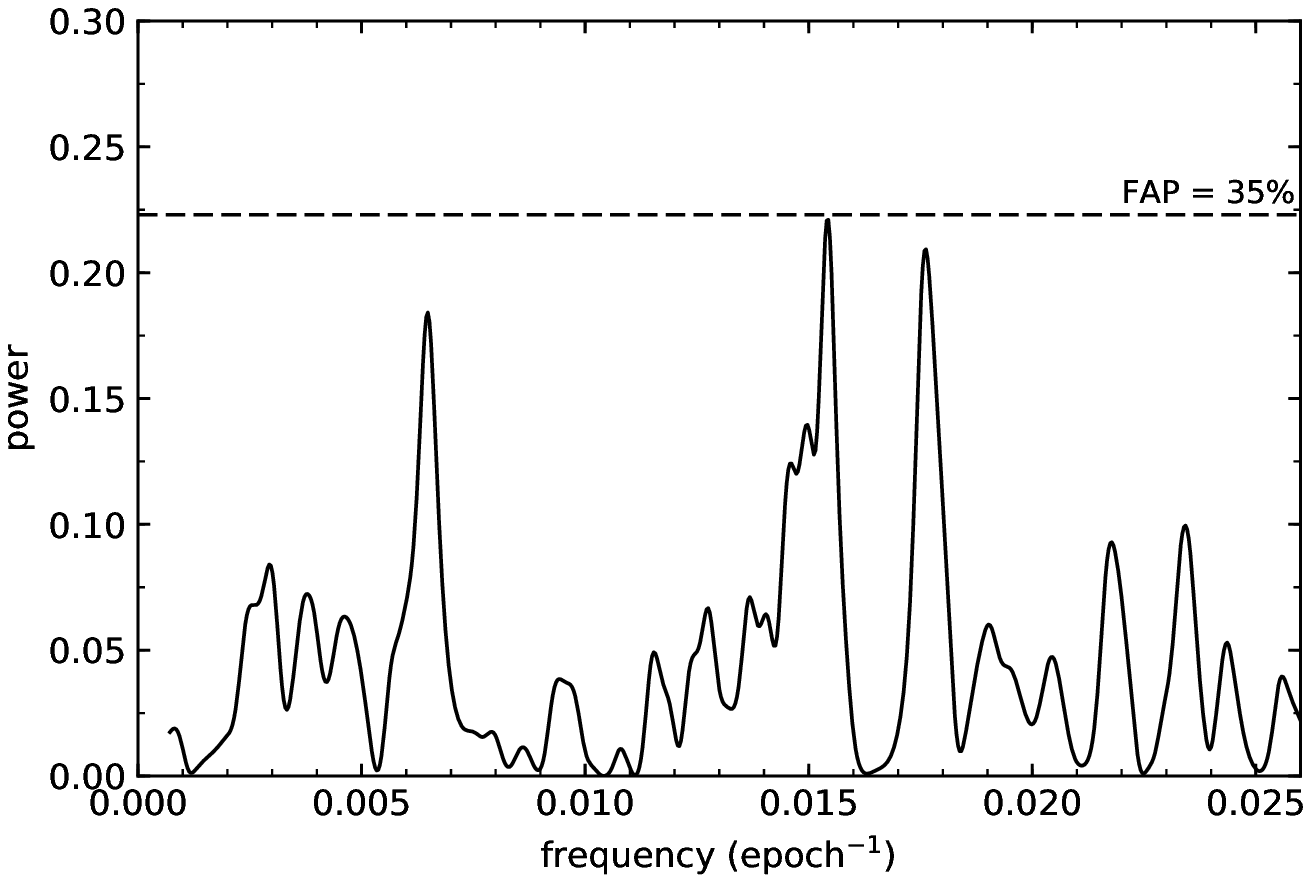}
\caption{The periodogram, i.e. the spectral power as a function of frequencies,
for all data points in $O-C$ diagram. 
The dash line indicates the false-alarm probability.}
\end{center}
\end{figure*}


\clearpage

\begin{thebibliography}{}

\bibitem[]{} Alonso, R., Brown, T. M., Torres, G., et al., 2004, ApJ, 613, L153

\bibitem[]{} Alsubai, K. A., Parley, N. R., Bramich, D. M., et al., 2011, MNRAS, 417, 709 

\bibitem[]{} Alsubai, K. A., Parley, N. R., Bramich, D. M., et al., 2013, Acta Astronomica, 63, 465

\bibitem[]{} Baglin, A., Auvergne, M., Boisnard, L., et al., 2006, 
in COSPAR Meeting, Vol. 36, 36$^{th}$ COSPAR Scientiﬁc Assembly

\bibitem[]{}
Baluev, R. V., Sokov, E. N., Jones, H. R. A., et al., 2019, MNRAS, 490, 1294 

\bibitem[]{} Blecic, J., Harrington, J., Madhusudhan, N., et al., 
2014, ApJ, 781, 116

\bibitem[]{} Borucki, W. J., Koch, D., Basri, G., et al., 
2010, Science, 327, 977

\bibitem[]{} Carter, J. A., Winn, J. N., 2009, ApJ, 704, 51 

\bibitem[]{} Claret, A., Bloemen, S., 2011, A\&A, 529, A75  

\bibitem[]{}
Chernov, S. V., Ivanov, P. B., Papaloizou, J. C. B., 2017,
MNRAS, 470, 2054

\bibitem[]{} Collins, K. A., Kielkopf, J. F.; Stassun, K. G., 2017, AJ, 153, 78

\bibitem[]{} Covino, E., Esposito, M., Barbieri, M., et al., 2013, A\&A, 554, A28  
\bibitem[]{} Dawson, R. I.,  Johnson, J. A., 2018, 
Annual Review of Astronomy and Astrophysics, 56, 175
%arXiv:1801.06117

\bibitem[]{} Eastman, J., Gaudi, B. S., Agol, E., 2013, PASP, 125, 83  

\bibitem[]{}
Foreman-Mackey, D., Hogg, D. W., Lang, D., Goodman, J., 2013, PASP, 125, 306

\bibitem[]{}
Fulton, B. J., Shporer, A., Winn, J. N., et al., 2011, AJ, 
142, 84   

\bibitem[]{} Gazak, J. Z., Johnson, J. A., Tonry, J., et al., 2012, 
Advances in Astronomy, 2012, 697967 

\bibitem[]{} Gayon, J., Bois, E., 2008, A\&A, 482, 665 

\bibitem[]{} Hellier, C., Anderson, D. R., Bouchy, F., et al., 2018, 
arXiv:1803.02224

\bibitem[]{} Howell, S. B., Sobeck, C., Haas, M., et al., 2014, PASP, 126, 398

\bibitem[]{} Hoyer, S., Palle, E., Dragomir, D., Murgas, F., 2016, 
AJ, 151, 137


\bibitem[]{} Ji, J., Li, G., Liu, L., 2002, ApJ, 572, 1041

\bibitem[]{} Jiang, I.-G., Ip, W.-H., 2001, A\&A, 367, 943

\bibitem{}
Jiang, I.-G., Ip, W.-H., Yeh, L.-C., 2003, ApJ, 582, 449

\bibitem{}  Jiang, I.-G., Yeh, L.-C., 2004a, AJ, 128, 923

\bibitem{}  Jiang, I.-G., Yeh, L.-C., 2004b,
Int. J. Bifurcation and Chaos, 14, 3153

\bibitem{}  Jiang, I.-G., Yeh, L.-C., 2007,
ApJ, 656, 534

\bibitem{}
Jiang, I.-G., Yeh, L.-C., Hung, W.-L., Yang, M.-S., 2006, MNRAS, 370,
1379

\bibitem{}
Jiang, I.-G., Yeh, L.-C., Chang, Y.-C., Hung, W.-L., 2007, AJ, 134,
2061

\bibitem{}
Jiang, I.-G., Yeh, L.-C., Chang, Y.-C., Hung, W.-L., 2009, AJ, 137,
329

\bibitem{}
Jiang, I.-G., Yeh, L.-C., Chang, Y.-C., Hung, W.-L., 2010, ApJS, 186,
48

\bibitem{} Jiang, I.-G., et al., 2013, AJ, 145, 68

\bibitem{}
Jiang, I.-G., Yeh, L.-C., Hung, W.-L., 2015,
MNRAS, 449, L65

\bibitem{} Jiang, I.-G., Lai, C.-Y., Savushkin, A., et al., 2016, AJ, 151, 17 

\bibitem{} Labadie-Bartz, J., Rodriguez, J. E., 
Stassun, K. G., et al., 2018, arXiv:1803.07559

\bibitem{} Lee, J. W., Youn, J.-H., Kim, S.-L., Lee, C.-U., Koo, J.-R., 
2011, PASJ, 63, 301

\bibitem[]{} Levrard, B., Winisdoerffer, C., Chabrier, G., 
2009, ApJ, 692, L9

\bibitem{} Linial, I., Gilbaum, S., Sari, Re{\'{\i}}em, 2018, ApJ, 860, 16 


\bibitem{} Maciejewski, G., Fernandez, M., Aceituno, F., et al., 
2018, Acta Astronomica, 68, 371

\bibitem{} Maciejewski, G., Dimitrov, D., Neuhauser, R., et al.,  
2010, MNRAS, 407, 2625

\bibitem{} Maciejewski, G., Dimitrov, D., Fernandez, M., et al., 2016,
A\&A, 588, L6 

\bibitem{} Maciejewski, G., Fern\'andez, M., Aceituno, F. J., et al., 2015, 
A\&A, 577, A109 

\bibitem{} Maciejewski, G., Niedzielski, A., Wolszczan, A., et al., 
2013, AJ, 146, 147  

\bibitem{} Mandel, K., Agol, E., 2002, ApJ, 580, L171 

\bibitem{} Mannaday, V. K., Thakur P., Jiang I.-G., et al., 2020, AJ, 160, 47 

\bibitem{} M{\"u}ller, H. M., Huber, K. F., Czesla, S., Wolter, U., Schmitt, J. H. M. M., 2013, A\&A, 560, A112 

\bibitem{} Murgas F., Pall\'{e} E., Zapatero Osorio M. R., et al., 2014, 
A\&A, 563, A41

\bibitem{} Patra, K. C., Winn, J. N., Holman, M. J., et al., 2017, 
AJ, 154, 4

\bibitem{} Pepper, J., Pogge, R. W., DePoy, D. L., et al., 2007, 
PASP, 119, 923

\bibitem{} Pollacco, D. L., Skillen, I., Collier Cameron, A., et al., 
2006, PASP, 118, 1407

%\bibitem{} Press, W. H., Rybicki, G. B., 1989, ApJ, 338, 277 

\bibitem{} Press, W. H., et al., 1992, Numerical Recipes in Fortran,
Cambridge University Press
  
\bibitem{} P{\"u}sk{\"u}ll{\"u} C., Soydugan, F., Erdem, A., Budding, E., 
2017, New Astronomy, 55, 39

\bibitem{} Tabachnik, S., Tremaine, S., 2002, MNRAS, 335, 151

\bibitem{} Talens, G. J. J., Spronck, J. F. P., Lesage, A.-L., et al., 
2017, A\&A, 601, A11

\bibitem{} Thakur, P., Mannadey, V. K., Jiang, I.-G., et al., 2018, 
Bulletin de la Société Royale des Sciences de Liège, 
in Proceedings of the First Belgo-Indian Network for 
Astronomy \& Astrophysics (BINA) workshop, November 2016, 
held in Nainital, India, Vol. 87, pp. 132-136

\bibitem{} von Essen C., Schr{\"o}ter S., Agol E., Schmitt J. H. M. M., 
2013, A\&A, 555, A92

\bibitem{} Wilkins, A. N., Delrez, L., Barker, A. J., et al., 2017,
ApJ, 836, L24 

\bibitem{} Winn, J. N., Holman, M. J., Carter, J. A., et al., 
2009, AJ, 137, 3826

\bibitem{}
Zechmeister, M., Kurster, M., 2009, A\&A, 496, 577

\bibitem{}
Zucker, S., Mazeh, T., 2002, ApJ, 568, L113


\end{thebibliography}
\end{document}